\documentclass[preprint,aps,12pt,preprintnumbers,eqsecnum,nofootinbib]{revtex4}
\usepackage{graphicx}
\usepackage{subfigure}

\usepackage{color}
\usepackage{amssymb,amsmath,amsfonts,bbold}
\usepackage{epstopdf} 
\usepackage{braket}

\newcommand{\beq}{\begin{equation}}
\newcommand{\enq}{\end{equation}}

\begin{document}
%
%  
% Title of paper
\title{\vspace*{0.5in} 
Composite gravity from a metric-independent theory of fermions 
\vskip 0.1in}
\author{Christopher D. Carone}\email[]{cdcaro@wm.edu}
\author{Joshua Erlich}\email[]{jxerli@wm.edu}

\affiliation{High Energy Theory Group, Department of Physics,
College of William and Mary, Williamsburg, VA 23187-8795}

\author{Diana Vaman}\email[]{dvaman@virginia.edu}

\affiliation{Department of Physics, University of Virginia,
Box 400714, Charlottesville, VA 22904, USA}

\date{\today}
\begin{abstract}
We present a metric-independent, diffeomorphism-invariant model with interacting fermions that contains a massless composite graviton in its 
spectrum.  The model is motivated by the supersymmetric D-brane action, modulated by a fermion potential.  The gravitational coupling is 
related to new physics at the cutoff scale that regularizes UV divergences. We also speculate on possible extensions of the model.
\end{abstract}
\pacs{}

\maketitle

\section{Introduction} \label{sec:intro}

In Refs.~\cite{Carone:2016tup,Carone:2017mdw}, we presented a generally covariant model of scalar fields in which gravity appears 
as an emergent phenomenon in the infrared.  The theory requires a physical cutoff that regularizes ultraviolet divergences, and the 
Planck scale is then related to the cutoff scale.   The existence of a massless graviton pole in a two-into-two scattering
amplitude was demonstrated nonperturbatively by a resummation of diagrams, that was made possible by working
in the limit of a large number of physical scalar fields $\phi^a$, for $a=1 \ldots N$.   Although that work assumed 
a particular organizing principle, namely that the theory could be obtained from a different starting point where an 
auxiliary metric field is eliminated by imposing a constraint that the energy-momentum tensor of the theory vanishes, 
one could have just as easily started directly with the resulting non-polynomial action
\begin{equation}
S=\int d^Dx\ \left(\frac{\tfrac D2-1}{V(\phi^a)} \right)^{\frac{D}{2}-1}
\sqrt {\bigg|\det \left(\sum_{a=1}^N \partial_\mu\phi^a \,\partial_\nu\phi^a 
+\sum_{I,J=0}^{D-1}\partial_\mu X^I \,\partial_\nu X^J\, \eta_{IJ}\right)\bigg|} \,\,\, ,
\label{eq:scalars}
\end{equation}
where $D$ is the number of spacetime dimensions and $V(\phi^a)$ is the scalar potential.  This action is reminiscent of a bosonic D-brane action~\cite{Polchinski:1996na}), modulated by the scalar potential $V(\phi^a)$. The $X^I$ fields will play the role of clocks and rulers after they are gauge-fixed, and provide a spacetime
backdrop for the theory.  The composite graviton in this theory couples at leading order
to the (non-vanishing) flat-space energy-momentum tensor of the physical scalars $\phi^a$;  graviton self-interactions are also generated, up 
to higher-derivative corrections~\cite{Carone:2017mdw}.  The vanishing of the Noether energy-momentum tensor of the model given by (\ref{eq:scalars}), which allows us to evade the Weinberg-Witten theorem~\cite{Weinberg:1980kq}, is a consequence of the fact that non-polynomial action in Eq.~(\ref{eq:scalars}) is manifestly metric independent and diffeomorphism invariant.\footnote{In string theory, there is an analogy with the Nambu-Goto action, which is metric-independent and diffeomorphism invariant. This leads to the Virasoro constraints, namely that the energy-momentum tensor is (classically) identically zero.}

If one were to take a theory like Eq.~(\ref{eq:scalars}) as a starting point for building more realistic theories with composite gravitons, two 
issues need to be addressed.  First, a compelling ultraviolet completion of the theory needs to be found.   
In Refs.~\cite{Carone:2016tup,Carone:2017mdw}, dimensional regulatization, with $D=4-\epsilon$ and $\epsilon$ taken to be finite, 
was used to provide a generally covariant cutoff.  The Planck scale was then determined in terms of the finite
value of $1/\epsilon$ and the dimensionful parameters of the theory.   A similar approach to composite gauge bosons can be found in Ref.~\cite{Suzuki:2016aqj}. The choice of dimensional regularization was motivated by convenience; alternatively, one could use a 
Pauli-Villars regulator, with the Planck scale fixed in terms of the Pauli-Villars scale, at the expense of 
complicating the relevant loop calculations.  In either case, the regulator is a placeholder for whatever generally 
covariant physics completes the theory in the ultraviolet.  In this paper we are agnostic about the ultraviolet completion, but there 
are various possibilities that are worthy of consideration. For example, it might be possible for theories 
of this type to be formulated covariantly on a discrete spacetime lattice, without going to a continuum limit.

The issue that we address in the present paper is whether similar models can be constructed
in the case of particles with higher spins.  We take the first step in this direction by presenting a model involving fermions fields 
whose action is similar to Eq.~(\ref{eq:scalars}), and showing that it leads to a massless, composite graviton state.   The fermion 
theory is described by a  metric-independent action which  is similar to that of a supersymmetric 
D-brane~\cite{Aganagic:1996nn} modulated by the fermion potential, though neither supersymmetry nor $\kappa$-symmetry are present (or required) in our model.

Our paper is organized as follows.  In the next section, we briefly summarize the approach used in 
Refs.~\cite{Carone:2016tup,Carone:2017mdw} to demonstrate that the scalar theory in Eq.~(\ref{eq:scalars}) includes a massless, 
composite spin-2 state.   In Sec.~\ref{sec:fmodel}, we present a model in which the graviton emerges as a bound state of fermions.   In Sec.~\ref{sec:discuss}, we discuss some subtleties of our analysis as well as possible generalizations of the model. In 
Sec.~\ref{sec:conc}, we summarize our conclusions.  We adopt the following conventions: the Minkowski metric has mostly minus 
signature, the Dirac matrices satisfy the Clifford algebra $\{\gamma^I,\,\gamma^J\}=2\, \eta^{IJ}$, and 
$\overline{\psi}=\psi^\dagger \gamma^0$ is the Dirac conjugate fermion.

%%%%%%%%%%%%%%%%%%%%%%%%%%%%%%%%%%%%
\section{Review of the scalar theory} \label{sec:review}

\begin{figure}[t]
  \begin{center}
    \includegraphics[width=.7\textwidth]{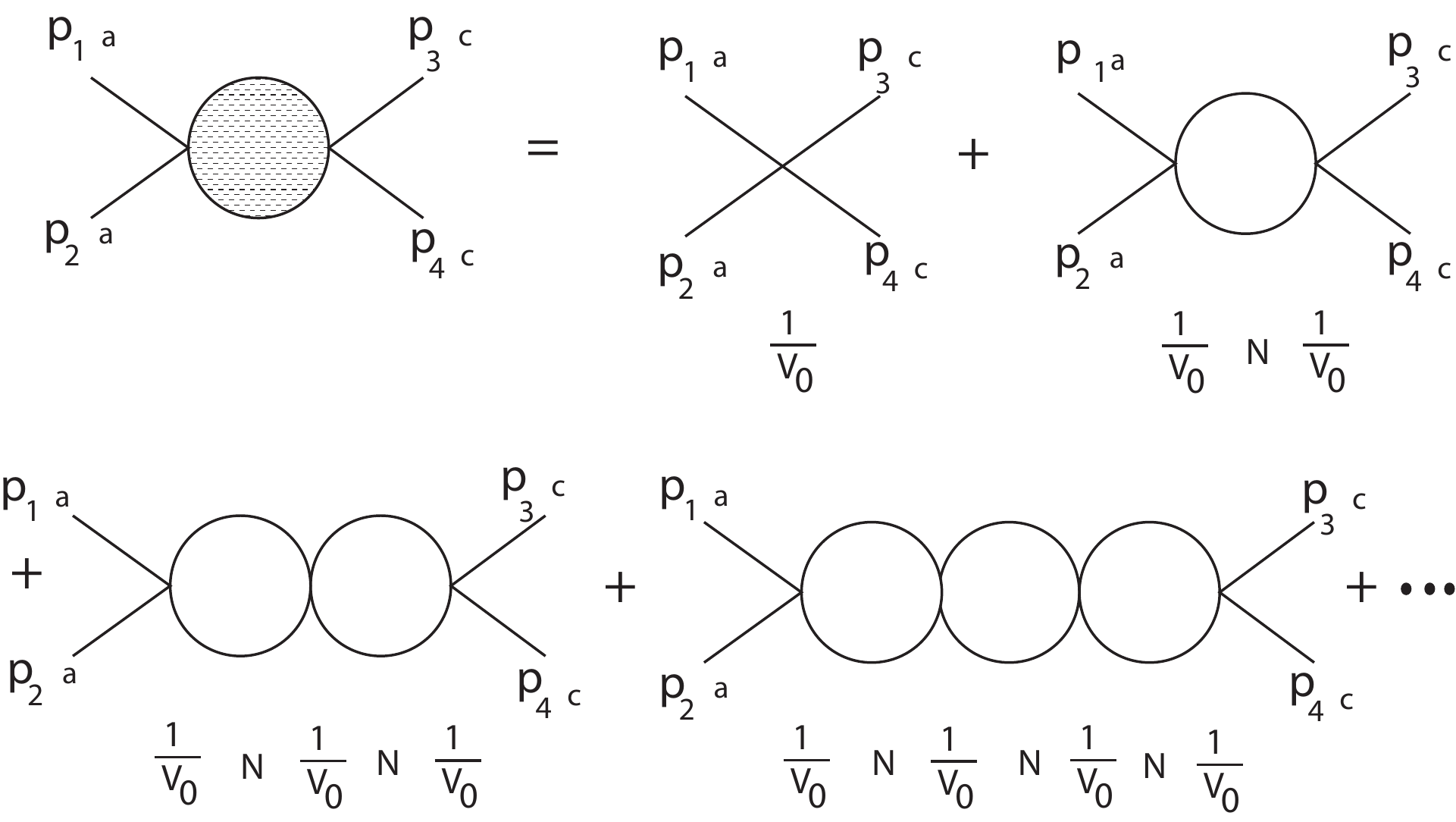}
    \caption{The scattering amplitude studied in Ref.~\cite{Carone:2016tup}}
        \label{fig:ampsc}
  \end{center}
\end{figure}

In this section, we briefly review the approach presented in Ref.~\cite{Carone:2016tup} to demonstrate the existence of a massless composite graviton in the scalar theory defined by Eq.~(\ref{eq:scalars}).   We first gauge-fix the general coordinate invariance of the theory by identifying the clock and ruler fields $X^I$ with the spacetime coordinates, up to a proportionality factor\footnote{The main assumption we make here, as in our previous work, is that the matrix $\partial_\mu X^I$ is non-singular, such that the static gauge (\ref{eq:staticgauge}) can be fixed.}
\begin{equation}
X^I=  x^\mu\delta_\mu^I \ \sqrt{\frac{V_0}{\tfrac{D}{2}-1}-c_1} , \ \ I=0,\dots,D-1 \, . \label{eq:staticgauge}
\end{equation}
Here, $V_0$ is the constant defined in the scalar potential by
\begin{equation}
V(\phi)=V_0+\Delta V(\phi^a)  \,\,\, ,
\label{eq:V0def}
\end{equation}
and we take $\Delta V(\phi^a)$ to represent the O($N$)-symmetric mass term
\begin{equation}
\Delta V(\phi^a)=\sum_{a=1}^N\frac{m^2}{2}\phi^a\phi^a-c_2 . \label{c_2} 
\end{equation}
The counterterms $c_1$ and $c_2$ will be explained below.  Eq.~(\ref{eq:V0def}) allows us to study the theory of 
Eq.~(\ref{eq:scalars}) perturbatively, via an expansion in powers of $1/V_0$.  We will also take the number of $\phi^a$ fields
to be large, $N \gg 1$, and work only to leading order in a $1/N$ expansion.  This makes it possible to perform a resummation 
of the leading-order diagrams that contribute to the two-into-two scattering amplitude shown in Fig.~\ref{fig:ampsc}.  The result of
this resummation allows the identification of a graviton pole, following a tuning of $V_0$ that is 
tantamount to the tuning of the cosmological constant.

Expanding the gauge-fixed action, one finds
\begin{eqnarray}
S=\int d^Dx&&\left\{\frac{V_0}{D/2-1}+\frac{1}{2}\sum_{a=1}^N \partial_\mu \phi^a \partial^\mu \phi^a -\Delta V(\phi^a )\right. \nonumber \\ &&
-\frac{\tfrac D2-1}{4V_0}\left[
\sum_{a=1}^N\partial_\mu\phi^a \partial_\nu\phi^a 
\,\sum_{b=1}^N\partial^\mu\phi^b\partial^\nu \phi^b
-\frac{1}{2}\left(\sum_{a=1}^N \partial_\mu \phi^a \partial^\mu \phi^a  \right)^{\!\!2\,}\right] \nonumber \\
&& \left.-\frac{\tfrac{D}{2}-1}{2}\frac{\Delta V(\phi^a)}{V_0}\sum_{a=1}^N \partial_\mu \phi^a\partial^\mu \phi^a+\frac{D}{4}\frac{(\Delta V(\phi^a))^2}{V_0}+{\cal O}\left(\frac{1}{V_0^2}\right)\right\} \,\,\, ,   \label{eq:Sexpansion}
\end{eqnarray}
where we have not displayed $c_1$ and $c_2$.   Their effects are simple to take into account, since these 
counterterms are chosen to normal order every occurrence of $\partial_\mu \phi^a \partial_\nu\phi^a$ and $\phi^a \phi^a$ in Eq.~(\ref{eq:Sexpansion}).   To be precise, any loop which can be formed by contracting the two $\phi^a$'s in $\partial_\mu \phi^a \partial_\nu\phi^a$ is set to zero by choice of the counterterm $c_1$, while any formed by contracting the two $\phi^a$'s in 
$m^2 \phi^a\phi^a$ is canceled by the counterterm $c_2$.  These choices reduce the number of relevant loop diagrams in 
the scattering calculation to those shown in Fig.~\ref{fig:ampsc}, at leading order in $1/N$.  The vertices in these diagrams originate
from the quartic scalar interactions of Eq.~(\ref{eq:Sexpansion}), which may be written more compactly as
\begin{equation}
{\cal L}_{int} = -\frac{1}{4 V_0} {\cal T}_{\mu\nu} \, \Pi^{\mu\nu|\alpha\beta}  \, {\cal T}_{\alpha\beta} \,\,\, ,
\label{eq:int1}
\end{equation}
where ${\cal T}_{\mu\nu}$ is the flat-space energy-momentum tensor
\begin{equation}
{\cal T}^{\mu\nu}= \sum_{a=1}^N\bigg[\partial^\mu \phi^a \partial^\nu \phi^a - \eta^{\mu\nu} \left(\frac{1}{2} \partial^\alpha \phi^a \partial_\alpha \phi^a - \frac{1}{2} m^2 \phi^a \phi^a \right)\bigg]  \,\,\, ,
\label{eq:tflat}
\end{equation}
and where 
\begin{equation}
 \Pi^{\alpha\beta|\lambda\kappa} \equiv \frac{1}{2}  \left[ (\tfrac D2-1)\left(\eta^{\alpha\lambda} \eta^{\beta\kappa} + \eta^{\alpha\kappa} \eta^{\beta\lambda}\right)
 - \eta^{\lambda\kappa}\eta^{\alpha\beta} \right] \,\,\, . \label{eq:Pi}
 \end{equation}
We may write the scattering amplitude of Fig.~\ref{fig:ampsc} in the form
\begin{equation}
i {\cal M} (p_1,a \,; p_2,a \rightarrow p_3,c\,; p_4,c)\equiv E_{\mu\nu}(p_1,p_2) [i \,A^{\mu\nu|\rho\sigma}(q) ] \,E_{\rho\sigma}(p_3,p_4)\,\,\, ,
\end{equation}
where $E_{\mu\nu}$ represents the Feynman rule for the external lines
\begin{equation}
E_{\mu\nu}(p_1,p_2) \equiv - (p_1^\mu \, p_2^\nu + p_1^\nu \, p_2^\mu) + \eta^{\mu\nu} (p_1 \cdot p_2 +m^2) \,\,\, ,
\label{eq:exln}
\end{equation}
and where $q=p_1+p_2=p_3+p_4$.   In evaluating the scattering amplitude, dimensional regularization is used as a regulator of the loop integrals.  It was shown in Ref.~\cite{Carone:2016tup}  that a massless, spin-two pole is obtained provided one chooses
\begin{equation}
V_0=-\frac{N(D/2-1)}{2} \frac{\Gamma(-D/2)}{(4\pi)^{D/2}} (m^2)^{D/2}    \,\, , \label{choice}
\end{equation}
a tuning that is necessary to achieve a vanishing cosmological constant.  One then finds
\begin{equation}
A^{\mu\nu|\rho\sigma}(q) =  -\frac{3 \,m^2}{D\, V_0}\, \left[(\tfrac D2-1) \,( \eta^{\nu\rho} \eta^{\mu\sigma} + \eta^{\nu\sigma} \eta^{\mu\rho}) 
- \eta^{\mu\nu} \eta^{\rho\sigma} \right] \, \frac{1}{q^2} +\cdots \,\,\, ,
\label{eq:ampsol}
\end{equation}
which can be compared to the expected one-graviton exchange amplitude in a free scalar theory
\begin{equation}
A^{\mu\nu|\rho\sigma}(q) = -\frac{M_{{\rm Pl}}^{2-D}}{D-2} \, \left[(\tfrac D2-1) \,( \eta^{\nu\rho} \eta^{\mu\sigma} + \eta^{\nu\sigma} \eta^{\mu\rho}) 
- \eta^{\mu\nu} \eta^{\rho\sigma} \right] \, \frac{1}{q^2} \,\,\, . \label{eq:onegraviton}
\end{equation}
Hence, one may identify the $D$-dimensional Planck mass $M_{{\rm Pl}}$ as 
\begin{equation}
M_{{\rm Pl}}= m\,\bigg[\frac{N \, \Gamma(1-\frac{D}{2})}{6\, (4 \pi)^{D/2} }\bigg]^{1/(D-2)} \,\,\, . \label{eq:mpscat}
\end{equation}
With $D=4-\epsilon$, positivity of the Planck mass requires the regulator $\epsilon$ to be small and negative.   

In the next section, we present a fermionic theory in which a massless, composite graviton is identified through a similar
large-$N$ diagrammatic resummation of a two-into-two scattering amplitude.  In Ref.~\cite{Carone:2016tup}, the resummation
of diagrams was achieved by solving a recursive formula that faithfully represents the complete set of 
diagrams shown in Fig.~\ref{fig:ampsc}.  The details of this approach will be revisited in the context of the fermionic model in the
next section.
 
%%%%%%%%%%%%%%%%%%%%%%%%%%%%%%%%%%%%%%%%
\section{A fermionic theory} \label{sec:fmodel}
%%%%%%%%%%%%%%%%%%%%%%%%%%%%%%%%%%%%%%%%%

Motivated by the scalar theory discussed in the previous section, we consider a similar metric-independent, reparametrization invariant action 
\begin{eqnarray}
S &=& - \int d^D x  \left(\frac{D/2-1}{|V(\psi)|}\right)^{D/2-1} \left[ \Big| \det\left(  \sum_{I,J}^{D-1} (\partial_\mu X^I + c_0 \,  {O^I}_\mu )   
(\partial_\nu X^J + c_0 \, {O^J}_\nu ) \, \eta_{IJ} \right)  \Big|\right]^{1/2} \nonumber\\
 &=& - \int d^D x \left(\frac{D/2-1}{|V(\psi)|}\right)^{D/2-1} \det\bigg(\partial_\mu X^I + c_0 {O^I}_\mu\bigg) ,\label{eq:fermaction}
\end{eqnarray}
where ${O^I}_\mu$ represents an operator that is bilinear in the fermion fields and the constant $c_0$ will be specified below.  We work with $N$ fermion flavors, in the large $N$ limit. We first consider 
the choice 
\begin{equation}
{O^I}_{\mu} \equiv \sum_{a=1}^N \frac{i}{2} \eta^{I J}\,\overline{\psi^a} \stackrel{\!\!\!\!\leftrightarrow}{\partial_{(\mu}} \gamma_{J)} \psi^a \,\,\, ,
\label{eq:symchoice}
\end{equation}
where $A\stackrel{\!\!\!\!\leftrightarrow}{\partial_{\mu}} B=A\partial_\mu B-(\partial_\mu A)B$. 
We use parentheses around a pair of indices to denote index symmetrization: $A_{(\mu \nu)}\equiv\frac{1}{2}\left(A_{\mu\nu}+A_{\nu\mu}\right)$, {\em etc}. In order to avoid notational clutter we will suppress the summation over the flavors from now on. We separate  the constant part of the potential into two parts, $V_0$ and $c_2$: 
\begin{equation}
V=V_0 + \Delta V-c_2 = V_0+:\!\Delta V\!:\,. 
\end{equation} The role of the constant $c_2$ is to normal-order $\Delta V$ such that \begin{equation}
\langle 0|\!:\!\Delta V\!: \!|0\rangle=0. \end{equation}
In the first line of Eq.~(\ref{eq:fermaction}), the signature of the vacuum spacetime metric appears explicitly in the factor of 
$\eta_{IJ}$; in the second line, the signature is determined by the Clifford algebra of the $\gamma$ matrices and the definition of
${O^I}_{\mu}$ in Eq.~(\ref{eq:symchoice}).

We further gauge-fix  the coordinate dependence of the clock and ruler fields as
\begin{equation}
X^I = \bigg(\sqrt{\frac{|V_0|}{(\frac{D}{2}-1)}} -c_1\bigg)\, x^\mu \, \delta_\mu^I, \,\,\,\,\,  I=0,\ldots,D-1 \,.
\end{equation}
This gauge choice is the same as in the scalar theory (with a  redefined $c_1$, but fulfilling the same purpose, namely  $-c_1\delta_\mu^I$ effectively normal-orders $c_0 \, O_\mu^I$) but allows $V_0$ to be of either sign.  Defining the normal-ordered flat-space matter energy-momentum tensor with vanishing vacuum expectation value,
\begin{equation}
t_{\mu\nu} \equiv : \!\frac{i}{2} \overline{\psi} \stackrel{\!\!\!\!\leftrightarrow}{\partial_{(\mu}} \gamma_{\nu )} \psi \!:- \eta_{\mu\nu} \left[ :\!\frac{i}{2} \overline{\psi} \stackrel{\!\!\!\leftrightarrow}{\partial_{\mu}} \gamma^{\mu} \psi \!:-:\! \Delta V \!:\right] \,\,\, ,
\end{equation}
and expanding in inverse powers of $V_0$, we find that the Lagrangian is given by
\begin{equation}
{\cal L} = \frac{2 V_0}{D-2} + (:\!O\!:-:\!\Delta V\!:) -\frac{D-2}{4 V_0} \left( t \cdot t - \frac{1}{D-1} \, t^2 \right) + \frac{D}{4(D-1) V_0} \Delta V^2 +{\cal O}(\frac{1}{V_0^2})\,\,\, ,
\label{eq:lag}
\end{equation}
where we introduced the notation $t \cdot t \equiv t_{\mu\nu} t^{\nu\mu}$, $t \equiv {t^\mu}_\mu$ and $O={O^\mu}_\mu$.   In this equation as well as those that follow we take $V_0<0$, which we find is necessary to assure the positivity of $M_P^2$. We assume, as in our scalar model, that $\Delta V$ includes only a mass term, 
\begin{equation}
\Delta V = m\, \overline{\psi} \, \psi \,\,\, ,
\end{equation}
so that the quantity $O-\Delta V$ represents the Lagrangian for a free Dirac fermion of mass $m$.  We want to stress that since every occurence  of $c_0 \, {O^I}_\mu$ comes with $-c_1\delta_\mu^I$ and every occurence of $\Delta V$ comes with $-c_2$, even higher-order interactions will be written in terms of the normal-ordered operators, $:\!{O^I}_\mu\!:$ and $:\!\Delta V\!:$. The value of $c_0$ has been fixed in Eq.~(\ref{eq:lag}) by the requirement that the fermion kinetic terms have canonical normalization, 
\begin{equation}
c_0 = \sqrt{D/2-1}\frac{\sqrt{|V_0|}}{V_0}   \,\,\,.
\end{equation}
Then, using dimensional regularization, we find 
\begin{equation}
c_1= 2\, N \sqrt{D/2-1}\frac{\sqrt{|V_0|}}{V_0}\Gamma(-D/2) \bigg(\frac{m^2}{4\pi}\bigg)^{D/2}\,\,\,.
\end{equation}

\begin{figure}[t]
  \begin{center}
    \includegraphics[width=.6\textwidth]{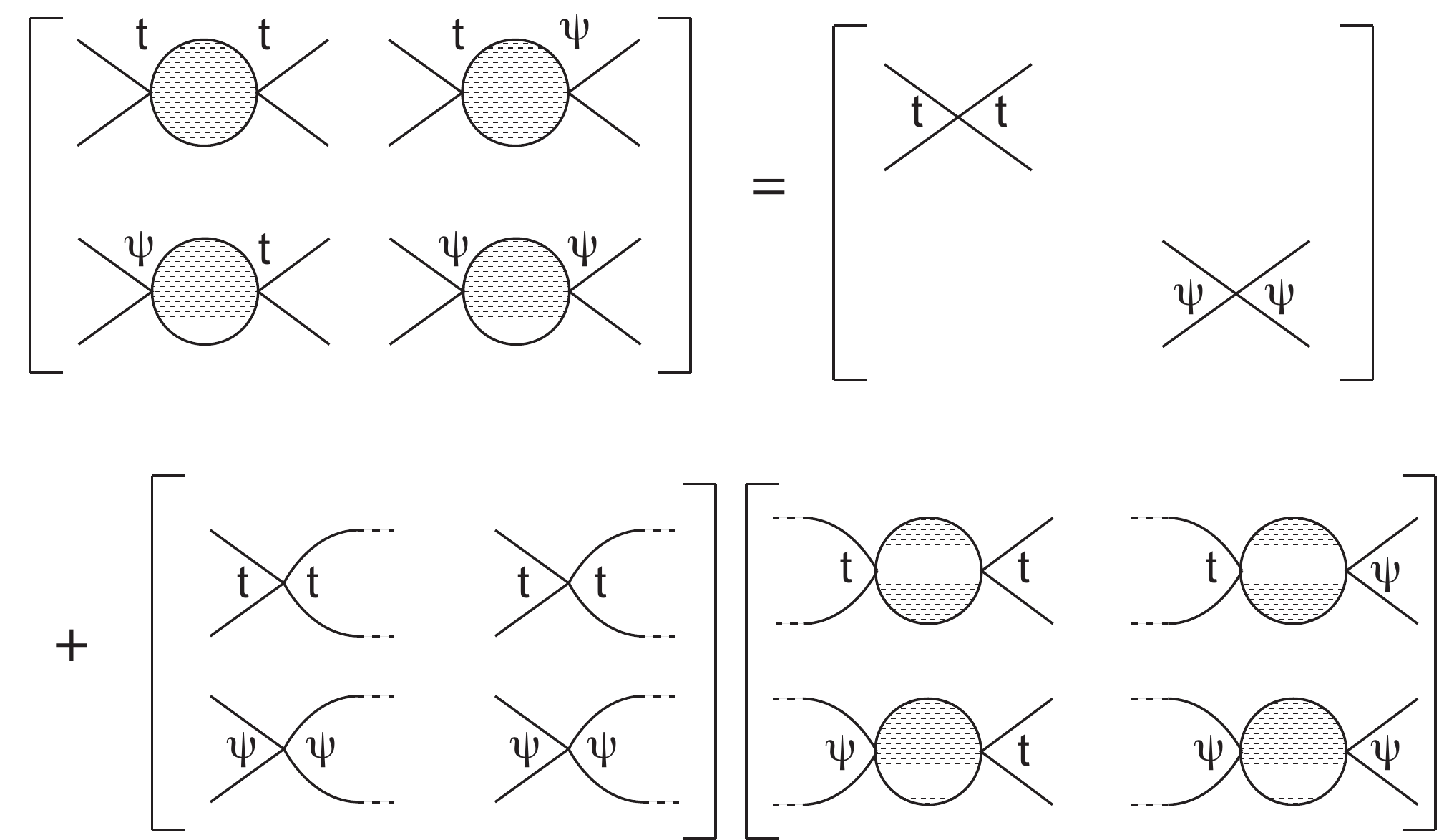}
    \caption{Diagrammatic representation of the recursion formula.}    
        \label{fig:blob}
  \end{center}
\end{figure}

From here on, we will be working with $D=4-\epsilon$ and to leading order in $\epsilon$.
The quartic interaction terms in Eq.~(\ref{eq:lag}) can be written
\begin{equation}
{\cal L}_{int} = - \frac{1}{4 V_0} \,t_{\mu\nu} \, \widetilde{\Pi}^{\mu\nu | \rho \sigma} \, t_{\rho \sigma} + \frac{1}{3} \frac{m^2}{V_0} (\overline{\psi} \, \psi )^2+{\cal O}(\frac{1}{V_0^2})
\,\,\, .
\label{eq:familiar}
\end{equation}
where
\begin{equation}
\widetilde{\Pi}^{\mu\nu | \rho \sigma} = 2\, \Pi^{\mu\nu | \rho \sigma} + \frac{1}{3} \, \eta^{\mu\nu} \eta^{\rho\sigma} \,\,\, . 
\end{equation}
Defining $\Psi_{\mu\nu} \equiv  \eta_{\mu\nu} \overline{\psi}\,\psi$, we may represent
the two types of interaction terms in a similar way
\begin{equation}
{\cal L}_{int} = - \frac{1}{4 V_0} \left[ t_{\mu\nu} \, \widetilde{\Pi}^{\mu\nu | \rho \sigma} \, t_{\rho \sigma} 
+\Psi_{\mu\nu}\, \widetilde{\Pi}^{'\mu\nu | \rho \sigma} \, \Psi_{\rho\sigma} \right] \,\,\, ,
\end{equation}
where  
\begin{equation}
\widetilde{\Pi}^{'\mu\nu | \rho \sigma}  = -\frac{1}{6} m^2 (\eta^{\mu\rho} \eta^{\nu \sigma} + \eta^{\mu\sigma} \eta^{\nu \rho})  \,\,\, .
\end{equation}
The operators $t_{\mu\nu}$ and $\Psi_{\mu\nu}$  lead to different Feynman rules for external lines, which we represent by
$E_{\mu\nu}$ and $E'_{\mu\nu}$, respectively.  With this notation, the full scattering amplitude is given by
\begin{equation}
i {\cal M} \equiv i \, \left(\begin{array}{cc} E_{\mu\nu} & E'_{\mu\nu}  \end{array}\right)
\left(\begin{array}{cc} A_{tt}^{\mu\nu|\rho\sigma} & A_{t\psi}^{\mu\nu|\rho\sigma}  \\
A_{\psi t}^{\mu\nu|\rho\sigma} & A_{\psi \psi}^{\mu\nu|\rho\sigma}  \end{array} \right) 
\left(\begin{array}{c} E_{\rho\sigma} \\ E'_{\rho\sigma} \end{array} \right) \, ,
\end{equation}
where
\begin{equation}
\left(\begin{array}{cc} A_{tt}^{\mu\nu|\rho\sigma} & A_{t\psi}^{\mu\nu|\rho\sigma} \\
A_{\psi t}^{\mu\nu|\rho\sigma} & A_{\psi \psi}^{\mu\nu|\rho\sigma} \end{array} \right)  = 
\left(\begin{array}{cc} A_{0\, tt}^{\mu\nu|\rho\sigma} & 0  \\
0 & A_{0 \, \psi \psi}^{\mu\nu|\rho\sigma} \end{array} \right)+
\left(\begin{array}{cc} K^{tt\,\mu\nu}_{\alpha\beta} & K^{t\psi\,\mu\nu}_{\alpha\beta} \\ 
K^{\psi t\,\mu\nu}_{\alpha\beta} & K^{\psi\psi\,\mu\nu}_{\alpha\beta} \end{array}\right) \left(\begin{array}{cc} A_{tt}^{\mu\nu|\rho\sigma} & A_{t\psi}^{\mu\nu|\rho\sigma} \\
A_{\psi t}^{\mu\nu|\rho\sigma} & A_{\psi \psi}^{\mu\nu|\rho\sigma} \end{array} \right) \, .
\label{eq:writtenout}
\end{equation}
The first term on the right-hand-side represents the tree-level contributions to the amplitude.  The four components of the kernel matrix each involve a 
one-loop calculation that differ by the choice of the two vertices which connect the loop to external lines on one end, and to the remaining scattering amplitude ``blob" on the other, as shown in  Fig.~\ref{fig:blob}.  
In these calculations, we retain terms only up to quadratic order in the momentum transfer $q$, which is sufficient for determining whether there is a pole at $q^2=0$.  Writing Eq.~(\ref{eq:writtenout}) as a two-by-two
matrix equation
\begin{equation}
\left[\delta^\mu_{(\alpha} \delta^\nu_{\beta)} \cdot \mathbb{1} - {K^{\mu\nu}}_{\alpha\beta}\right] A^{\alpha \beta | \rho \sigma} = A_0^{\mu\nu | \rho\sigma}
\,\,\, ,
\end{equation}
we need to invert the two-by-two matrix of tensors in square brackets on the left-hand-side, up to terms that vanish when contracted
with the external line factor $E_{\lambda\kappa}$.   For $D=4-\epsilon$ dimensions, and with a tuning analogous to the one we encountered in the scalar theory,
\begin{equation}
\frac{N m^4}{8 \pi^2 V_0 \epsilon} = 1 \,\,\, ,
\label{eq:tune}
\end{equation}
we find that the matrix $\delta^\mu_{(\alpha} \delta^\nu_{\beta)} \cdot \mathbb{1} - {K^{\mu\nu}}_{\alpha\beta}$ has the 
following four components:
\begin{eqnarray}
&& \delta^\mu_{(\alpha} \delta^\nu_{\beta)} -{{K_{tt}}^{\mu\nu}}_{\alpha\beta}  = - \frac{1}{3} \frac{q^2}{m^2}\,\delta^\nu_{(\alpha} \delta^\mu_{\beta)}
+ \frac{1}{9 m^2} \eta^{\mu\nu} (q^2 \eta_{\alpha\beta} -q_\alpha q_\beta)
\nonumber \\ &&+\frac{1}{6 m^2}(
 q^\mu \delta_\alpha^\nu q_\beta
+ q^\mu \delta^\nu_\beta q_\alpha
+ q^\nu \delta_\alpha^\mu q_\beta
+ q^\nu \delta_\beta^\mu q_\alpha ) -\frac{1}{3 m^2} q^\mu q^\nu \eta_{\alpha\beta}   \,\,\,,
\label{eq:kerntt}
\end{eqnarray}
\begin{equation}
 - {{K_{t\psi}}^{\mu\nu}}_{\alpha\beta} = - \frac{2}{3 \, m^3} q^\mu q^\nu \eta_{\alpha\beta} \,\,\, ,
\label{eq:kerntp}
\end{equation}
\begin{equation}
-{{K_{\psi t}}^{\mu\nu}}_{\alpha\beta} =  - \frac{1}{9 \, m} \eta^{\mu\nu} (q^2 \eta_{\alpha\beta} -q_\alpha q_\beta) \,\,\, ,
\label{eq:kernpt}
\end{equation}
\begin{equation}
\delta^\mu_{(\alpha} \delta^\nu_{\beta)}- {{K_{\psi\psi}}^{\mu\nu}}_{\alpha\beta} = \delta^\mu_{(\alpha} \delta^\nu_{\beta)} + 2 \,\eta^{\mu\nu} \eta_{\alpha\beta} \, (1-\frac{1}{6} \frac{q^2}{m^2}) \,\,\,\,.
\label{eq:kernpp}
\end{equation}
Parameterizing the components of the inverse matrix
\begin{equation}
\left(\begin{array}{cc} {{\cal W}^{\lambda\kappa}}_{\mu\nu} & {{\cal X}^{\lambda\kappa}}_{\mu\nu} \\
{{\cal Y}^{\lambda\kappa}}_{\mu\nu}& {{{\cal Z}}^{\lambda\kappa}}_{\mu\nu} \end{array} \right) 
\left(\begin{array}{cc} \delta^\mu_{(\alpha} \delta^\nu_{\beta)}-{{K_{tt}}^{\mu\nu}}_{\alpha\beta} & -{{K_{t\psi}}^{\mu\nu}}_{\alpha\beta} \\
-{{K_{\psi t}}^{\mu\nu}}_{\alpha\beta} & \delta^\mu_{(\alpha} \delta^\nu_{\beta)}-{{K_{\psi\psi}}^{\mu\nu}}_{\alpha\beta} \end{array}\right)
= \left(\begin{array}{cc} \delta^\lambda_{(\alpha} \delta^\kappa_{\beta)} & 0 \\ 0 &  \delta^\lambda_{(\alpha} \delta^\kappa_{\beta)} 
\end{array}\right) \,\,\, , \label{eq:inversematrix}
\end{equation}
again, up to terms on the right-hand-side that vanish when acting on the external line factor $E_{\lambda\kappa}$, 
we find
\begin{equation}
{{\cal W}^{\lambda\kappa}}_{\mu\nu} = -\frac{3 \, m^2}{q^2} \, \delta^\lambda_{(\mu} \delta^\kappa_{\nu)}
+\left(\frac{2}{9}-w -\frac{3}{2}\frac{m^2}{q^2}\right) \, \eta^{\lambda\kappa} \eta_{\mu\nu} + w \, \eta^{\lambda\kappa} \frac{q_\mu q_\nu}{q^2} \,\,\, ,
\label{eq:W}
\end{equation}
\begin{equation}
{{\cal X}^{\lambda\kappa}}_{\mu\nu} = -\frac{1}{9 \, m} \, \eta^{\lambda\kappa} \eta_{\mu\nu} \,\,\, ,
\label{eq:X}
\end{equation}
\begin{equation}
{{\cal Y}^{\lambda\kappa}}_{\mu\nu} = \left( -\frac{m}{18} -y \right)\,\eta^{\lambda\kappa}\, \eta_{\mu\nu} + y \,\eta^{\lambda\kappa} \frac{q_\mu q_\nu}{q^2} 
\,\,\,,
\label{eq:Y}
\end{equation}
\begin{equation}
 {{{\cal Z}}^{\lambda\kappa}}_{\mu\nu} = \delta^\lambda_{(\mu} \delta^\kappa_{\nu)} - \frac{2}{9} \, \eta^{\lambda\kappa}\, \eta_{\mu\nu} \,\,\,.
 \label{eq:Z}
 \end{equation}
In these expressions, the constants $w$ and $y$ are arbitrary; we will see that the pole part of the scattering amplitude will be independent of their values.   

The solution for the amplitude is given by   
\begin{equation}
i {\cal M} = i \,
\left(\begin{array}{cc} E_{\lambda\kappa} & E'_{\lambda\kappa}  \end{array}\right) 
\left(\begin{array}{cc} {{\cal W}^{\lambda\kappa}}_{\mu\nu} & {{\cal X}^{\lambda\kappa}}_{\mu\nu} \\
{{\cal Y}^{\lambda\kappa}}_{\mu\nu}& {{{\cal Z}}^{\lambda\kappa}}_{\mu\nu} \end{array} \right) 
\left(\begin{array}{cc} A_{0\, tt}^{\mu\nu|\rho\sigma} & 0  \\
0 & A_{0 \, \psi \psi}^{\mu\nu|\rho\sigma} \end{array} \right) \left(\begin{array}{c} E_{\rho\sigma} \\ E'_{\rho\sigma} \end{array} \right) \, ,
\end{equation}
or more explicitly
\begin{eqnarray}
{\cal M} &=& -\frac{1}{2 V_0} \left[ E_{\lambda\kappa} {{\cal W}^{\lambda\kappa}}_{\mu\nu} \widetilde{\Pi}^{\mu\nu|\rho\sigma} E_{\rho\sigma} 
 + E_{\lambda\kappa} {{\cal X}^{\lambda\kappa}}_{\mu\nu} \widetilde{\Pi}^{'\mu\nu|\rho\sigma} E'_{\rho\sigma} \right. \nonumber \\
&&\left.+ E'_{\lambda\kappa} {{\cal Y}^{\lambda\kappa}}_{\mu\nu} \widetilde{\Pi}^{\mu\nu|\rho\sigma} E_{\rho\sigma}
+ E'_{\lambda\kappa} {{\cal Z}^{\lambda\kappa}}_{\mu\nu} \widetilde{\Pi}^{'\mu\nu|\rho\sigma} E'_{\rho\sigma}
\right]
\label{eq:finalM}
\end{eqnarray}
It is clear from Eqs.~(\ref{eq:X}) and (\ref{eq:Z}) that the second and fourth terms in Eq.~(\ref{eq:finalM}) have no $1/q^2$ poles.   Eq.~(\ref{eq:Y}) only provides a $1/q^2$ via the $q^\mu q^\nu /q^2$ term; however this term 
only contributes to the scattering amplitude via contractions with an $\eta_{\mu\nu}$ in Eq.~(\ref{eq:finalM}), so the third term in that expression is also free of $1/q^2$ poles.   The only possibility of a surviving pole is in the term proportional to ${{\cal W}^{\lambda\kappa}}_{\mu\nu} \widetilde{\Pi}^{\mu\nu|\rho\sigma}$ (the part of the amplitude that connects two flat-space energy-momentum tensors); using our result in Eq.~(\ref{eq:W}) we find
\begin{equation}
{{\cal W}^{\lambda\kappa}}_{\mu\nu} \widetilde{\Pi}^{\mu \nu | \rho\sigma} = - 6 \, \frac{m^2}{q^2} \,\Pi^{\lambda\kappa | \rho\sigma} \,\,\, ,
\end{equation}
which is independent of the constant $w$, and is proportional to the correct tensor structure for the spin-$2$ graviton propagator.  Taking into account the coefficient of the tree-level amplitude as well as the fine-tuning condition, the pole-part of the amplitude may be written
\begin{equation}
{A^{\lambda\kappa | \rho\sigma}} = \frac{24 \pi^2 \epsilon}{m^2 N} \, \frac{{\Pi^{\lambda\kappa | \rho\sigma}}}{q^2} \,\,\, .
\end{equation}
With $\epsilon<0$ [and $V_0<0$ from Eq.~(\ref{eq:tune})], we identify the Planck mass defined in Eq.~(\ref{eq:onegraviton}) as
\begin{equation}
M_P^2 = \frac{m^2 N}{24 \pi^2 |\epsilon|} \,\,\, .
\end{equation}
It is not surprising that the sign of $V_0$ is opposite to what one encounters in the scalar theory of Ref.~\cite{Carone:2016tup}, 
due to the additional minus sign that originates from the fermion loop.

Finally, we note that our choice of symmetric indices in Eq.~(\ref{eq:symchoice}) allows us to illustrate our result in
the simplest way possible.   Had we assumed no symmetry of indices, we would decompose ${O_I}_\mu$ into 
symmetric and antisymmetric parts.  This leads to a new quartic fermion interaction that is quadratic in the antisymmetric part
of ${O_I}_\mu$ by itself, and corresponds to a distinct external line Feynman rule.  We would then have to consider a more 
cumbersome three-by-three matrix kernel analysis.  However, we would find in the end that the graviton pole arises from iteration of 
loops involving the other interactions, a point that is demonstrated more directly with the model presented in this section.

%%%%%%%%%%%%%%%%%%%%%%%%%%%%%%%%%%%%%%%%%%
\section{Discussion}\label{sec:discuss}
%%%%%%%%%%%%%%%%%%%%%%%%%%%%%%%%%%%%%%%%%%%

It is likely that the fundamental theory of nature will incorporate diffeomorphism invariance in its description. It is also possible that the fundamental theory will serve to regularize the divergences of quantum field theory, as happens in string theory. In a semiclassical setting, Sakharov explained that the assumption of a diffeomorphism-invariant regulator in a  quantum field theory is almost guaranteed to give rise to the Einstein-Hilbert term in the effective action \cite{Sakharov:1967pk}. This has been confirmed in a number of toy models by direct computation \cite{Akama:1977hr,Visser:2002ew}. However, subsequent quantization of the spacetime metric generally leaves the basic puzzles of quantum gravity, such as the problem of time \cite{DeWitt:1967yk,Isham:1992ms}, unaddressed.

The existence of scalar fields that play the role of clock and rulers would resolve the problem of time and allow for an emergent geometric 
description of spacetime. In this setting there is a natural expansion of the emergent spacetime about Minkowski space (or about any 
other spacetime after suitable modification of the theory~\cite{Chaurasia:2017ufl}), and as we have seen there exists a perturbative 
expansion of scattering amplitudes that demonstrates the existence of an emergent gravitational interaction and a massless composite 
spin-2 state in the spectrum.  However, this approach presents some subtleties that we now mention.  With the fine-tuned choice for the 
potential $V_0$, we find that the leading order term in $1/\epsilon$ in our gauge-fixing condition 
$X^I=(\sqrt{|V_0|/(D/2-1)} - c_1)\, x^\mu \, \delta_{\mu}^I$ vanishes; the same observation was made in the context of the 
purely scalar theory in Ref.~\cite{Carone:2017mdw}.  However, in order for the 
gauge fixed choice for the $X^I$ to be related to a class of nontrivial field configurations by diffeomorphism transformations, we must keep 
the gauge-fixed profile of the $X^I$ such that $\partial_\mu X^I$ is a non-singular matrix.  This suggests that a way to define the theory is 
via the limit as $V_0$ approaches the fine-tuned value.   The discussion in Secs.~\ref{sec:review} and \ref{sec:fmodel} establishes the 
massless graviton state by working at the limit point rather than taking the limit.   

Another issue is that the theory described by the action in Eq.~(\ref{eq:fermaction}) is diffeomorphism invariant, but 
not locally Lorentz invariant, contrary to what one might expect for a theory of fermions that have gravitational interactions.  Nevertheless, 
other quantum theories of gravity with global rather than local Lorentz invariance have been proposed, taking into account that observational constraints are only sensitive to certain local-Lorentz-violating interactions that may be generated at
higher order in the effective action~\cite{Wetterich:2003wr}. Those models, as well as the one presented in 
Sec.~\ref{sec:fmodel}, illustrate that the existence of a massless spin-2 graviton in the spectrum follows only from 
the diffeomorphism invariance of the theory.  Rather than focus here on phenomenological issues associated with the 
absence of local Lorentz invariance, we instead outline how it may be incorporated in the theory:  Specifically, start with an action describing the clock and ruler fields and the $N$ massive free fermions coupled with an auxilliary vielbein $e_\mu^I$
\begin{equation}
S=\int d^D x \det({e_\mu^M} )\bigg(\frac 12 e^\nu_J e^\rho_K \eta^{JK} \eta_{IL}\partial_\nu X^I \partial_\rho X^L + i e^\nu_J\bar \psi \gamma^J D_\nu \psi - (m\bar\psi \psi+V_0-c_2)\bigg)\label{eq:aux0}
\end{equation}
where the covariant derivative $D_\mu$ includes the spin connection  term which ensures that the fermions in (\ref{eq:aux0}) transform as spinors under local Lorentz transformations. The spin connection is defined in terms of the auxiliary vielbein and its derivatives as usual.
One can impose once more the vanishing energy-momentum tensor constraint, and solve for the auxiliary vielbein $e_\mu^I$ order by 
order in $1/V_0$.  The resulting metric-independent theory would have both diffeomorphism invariance and local Lorentz symmetry.   

It is also interesting to note that our gauge-fixed model bears a striking resemblance to a supersymmetric D-brane 
action (compare for example with Eq.~(1) in Ref.~\cite{Aganagic:1996nn}). The only significant difference (besides having a large number of fermion flavors) is that we are interested in massive fermions\footnote{We recall that the Planck scale is proportional to the fermion mass in this approach.}, and so we added a modulating fermion potential to the action (we cannot add a potential term to the square-root DBI-like term, because in doing so we would be losing diffeomophism invariance which is crucial for our emergent gravity class of models; so the only option to add the potential is by mutiplication with the DBI-like term). Given that the D-brane action offered a template for emergent gravity for scalars and fermions alike, we offer some comments on coupling the spin-0 and spin-1/2 fields to a gauge field on the brane. In the abelian case one conjecture is to modify the action (\ref{eq:fermaction}) by adding a term proportional to the field strength, as in the DBI action~\cite{Polchinski:1996na}:
\begin{eqnarray}
S = \int d^D x &&\left(\frac{D/2-1}{V(\psi)}\right)^{D/2-1} \left[ \Big| \det\left(  \sum_{I,J}^{D-1} \left(\partial_\mu X^I + c_0 \,  i \overline{\psi}\gamma^I (\partial_\mu-i A_\mu)\psi\right) \right.\right.  \nonumber \\
&&\left. \left(\partial_\nu X^J + c_0 \,i \overline{\psi}\gamma^J (\partial_\nu-i A_\nu)\psi\right) \right) \, \eta_{IJ}   +2\pi\beta^\prime F_{\mu\nu} \Big|\Bigg]^{1/2}, \label{eq:conj1abelian}
\end{eqnarray}
where $\beta^\prime$ is a parameter ultimately related to the gauge coupling, and $F_{\mu\nu}=\partial_\mu A_\nu-\partial_\nu A_\mu$.  In (\ref{eq:conj1abelian}), we also replace partial derivatives with gauge-covariant derivatives. This takes us away from the supersymmetric D-brane action, where the fermions are not charged~\cite{Aganagic:1996nn}. For non-abelian supersymmetric D-branes, the fermions belong to the adjoint representation of the gauge group, as required by supersymmetry. Our model on the other hand only requires diffeomorphism and gauge invariance and it is not bound by supersymmetry considerations, which accounts for the departure from the supersymmetric D-brane actions.

 The definition of the theory by way of the auxiliary vielbein also suggests an extension that includes non-Abelian gauge fields.  We add to the auxiliary field action (\ref{eq:aux0}) the curved-space gauge-field kinetic term and interactions. If we then formally replace the vielbein by the solution to the constraint of vanishing energy-momentum tensor ({\em i.e.} the vielbein equation of motion) as a series expansion in $1/V_0$, we are led to a theory containing dynamical gauge fields and gauge-coupled fermions. By virtue of diffeomorphism invariance we expect the regulated theory to also provide an emergent gravitational interaction. We do not follow through with this approach here, although this would seem to be a direct path to a theory of emergent gravity coupled to the standard model. The possibility that {\em all} gauge and gravitational interactions are emergent should also be considered, motivated for example by the model of Refs.~\cite{Amati:1981rf,Amati:1981tu}, but with the addition of clock and ruler fields to allow the gauge fixing of diffeomorphism invariance and a perturbative analysis as in this paper.  

%%%%%%%%%%%%%%%%%%%%%%%%%%%%%%%%%%%%%%%%%%%%%%%%%%%%%%
\section{Conclusions}\label{sec:conc}
%%%%%%%%%%%%%%%%%%%%%%%%%%%%%%%%%%%%%%%%%%%%%%%%%%%%%%%%
We have demonstrated the existence of a long-range gravitational interaction consistent with general relativity in a metric-independent but diffeomorphism-invariant theory with spin-1/2 fermions. The model was motivated by the action describing spinors on a 
D-brane~\cite{Aganagic:1996nn}. The strength of the emergent gravitational interaction depends on the ultraviolet regulator for the theory; here we have used dimensional regularization as a proxy for the physical regulator that would provide a complete definition of the theory. Clock and ruler fields provide the spacetime backdrop of the theory, and give rise to a natural expansion of the emergent spacetime about Minkowski space after a parameter tuning that is tantamount to setting the cosmological
constant to zero~\cite{Carone:2016tup}. 

The approach to quantum gravity that we have explored here and in 
Refs.~\cite{Carone:2016tup,Carone:2017mdw} and \cite{Chaurasia:2017ufl} suggests that a long-range gravitational interaction consistent with general relativity is straightforward to arrange, as long as a physical covariant regulator exists.   In order to maintain the massless composite graviton in the spectrum of the effective low-energy theory, the physical regulator must preserve the diffeomorphism invariance of the effective theory. Insight into possible physical regulators can be found in other approaches to quantum gravity, even though here gravity is emergent rather than quantized directly. For example, if spacetime is discrete as in the causal dynamical triangulations approach \cite{Ambjorn:2012jv}, then a scale associated with that discreteness could be associated with the regulator scale; alternatively, more radical modifications of quantum theory at short distances could be responsible for regularizing field-theory divergences \cite{Markopoulou:2003ps,Erlich:2018qfc} and the resulting emergent gravitational interaction. Although here we analyzed a toy model in a large-$N$ limit including only fermions and the clock and ruler fields, it should be possible to extend these results in a more general setting, including one in which the low-energy effective description is that of the standard model coupled to Einstein gravity.

%%%%%%%%%%%%%%%%%%%%%%%%%%%%%%%%%%%%%%%%%%%%%%%%%%%%%%%%%%%
\begin{acknowledgments}  
The work of C.D.C. and J.E. was supported by the NSF under Grant PHY-1819575.  The work of DV was supported in part by 
the DOE grant DE-SC0007894.
\end{acknowledgments}
%%%%%%%%%%%%%%%%%%%%%%%%%%%%%%%%%%%%%%%%%%%%%%%%%%%%%%%%%%%     

%\appendix
%\section{} \label{sec:appendix}


\begin{thebibliography}{99}

\bibitem{Carone:2016tup} 
  C.~D.~Carone, J.~Erlich and D.~Vaman,
  ``Emergent Gravity from Vanishing Energy-Momentum Tensor,''
  JHEP {\bf 1703}, 134 (2017)
%  doi:10.1007/JHEP03(2017)134
  [arXiv:1610.08521 [hep-th]].
  %%CITATION = doi:10.1007/JHEP03(2017)134;%%
  
 \bibitem{Carone:2017mdw} 
  C.~D.~Carone, T.~V.~B.~Claringbold and D.~Vaman,
  ``Composite graviton self-interactions in a model of emergent gravity,''
  Phys.\ Rev.\ D {\bf 98}, no. 2, 024041 (2018)
%  doi:10.1103/PhysRevD.98.024041
  [arXiv:1710.09367 [hep-th]].
  %%CITATION = doi:10.1103/PhysRevD.98.024041;%% 

\bibitem{Polchinski:1996na} 
  J.~Polchinski,
  ``Tasi lectures on D-branes,''
  hep-th/9611050.
  %%CITATION = HEP-TH/9611050;%%
  %1240 citations counted in INSPIRE as of 12 Dec 2018

\bibitem{Weinberg:1980kq} 
  S.~Weinberg and E.~Witten,
  ``Limits on Massless Particles,''
  Phys.\ Lett.\  {\bf 96B}, 59 (1980).
%  doi:10.1016/0370-2693(80)90212-9
  %%CITATION = doi:10.1016/0370-2693(80)90212-9;%%
  
\bibitem{Suzuki:2016aqj} 
  M.~Suzuki,
  ``Composite gauge-bosons made of fermions,''
  Phys.\ Rev.\ D {\bf 94}, no. 2, 025010 (2016)
%  doi:10.1103/PhysRevD.94.025010
  [arXiv:1603.07670 [hep-th]].
  %%CITATION = doi:10.1103/PhysRevD.94.025010;%%  
  
%\cite{Aganagic:1996nn}
\bibitem{Aganagic:1996nn} 
  M.~Aganagic, C.~Popescu and J.~H.~Schwarz,
  ``Gauge invariant and gauge fixed D-brane actions,''
  Nucl.\ Phys.\ B {\bf 495}, 99 (1997)
  % doi:10.1016/S0550-3213(97)00180-6
  [hep-th/9612080].
  %%CITATION = doi:10.1016/S0550-3213(97)00180-6;%%
  %338 citations counted in INSPIRE as of 06 Dec 2018
  
  \bibitem{Sakharov:1967pk} 
  A.~D.~Sakharov,
  ``Vacuum quantum fluctuations in curved space and the theory of gravitation,''
  Sov.\ Phys.\ Dokl.\  {\bf 12}, 1040 (1968)
  [Dokl.\ Akad.\ Nauk Ser.\ Fiz.\  {\bf 177}, 70 (1967)]
  [Sov.\ Phys.\ Usp.\  {\bf 34}, no. 5, 394 (1991)]
  [Gen.\ Rel.\ Grav.\  {\bf 32}, 365 (2000)]
  [Usp.\ Fiz.\ Nauk {\bf 161}, no. 5, 64 (1991)].
%  doi:10.1070/PU1991v034n05ABEH002498
  %%CITATION = doi:10.1070/PU1991v034n05ABEH002498;%%
  
  \bibitem{Akama:1977hr} 
  K.~Akama, Y.~Chikashige, T.~Matsuki and H.~Terazawa,
  ``Gravity and Electromagnetism as Collective Phenomena: A Derivation of Einstein's General Relativity,''
  Prog.\ Theor.\ Phys.\  {\bf 60}, 868 (1978).
 % doi:10.1143/PTP.60.868
  %%CITATION = doi:10.1143/PTP.60.868;%%
  
  \bibitem{Visser:2002ew} 
  M.~Visser,
  ``Sakharov's induced gravity: A Modern perspective,''
  Mod.\ Phys.\ Lett.\ A {\bf 17}, 977 (2002)
%  doi:10.1142/S0217732302006886
  [gr-qc/0204062].
  %%CITATION = doi:10.1142/S0217732302006886;%%
  %135 citations counted in INSPIRE as of 29 Nov 2018
  
  \bibitem{DeWitt:1967yk} 
  B.~S.~DeWitt,
  ``Quantum Theory of Gravity. 1. The Canonical Theory,''
  Phys.\ Rev.\  {\bf 160}, 1113 (1967).
 % doi:10.1103/PhysRev.160.1113
  %%CITATION = doi:10.1103/PhysRev.160.1113;%%
  
  \bibitem{Isham:1992ms} 
  C.~J.~Isham,
  ``Canonical quantum gravity and the problem of time,''
  NATO Sci.\ Ser.\ C {\bf 409}, 157 (1993)
  [gr-qc/9210011].
  %%CITATION = GR-QC/9210011;%%

  \bibitem{Chaurasia:2017ufl} 
  S.~Chaurasia, J.~Erlich and Y.~Zhou,
  ``Curved Backgrounds in Emergent Gravity,''
  Class.\ Quant.\ Grav.\  {\bf 35}, no. 11, 115008 (2018)
%  doi:10.1088/1361-6382/aabac0
  [arXiv:1710.07262 [hep-th]].
  %%CITATION = doi:10.1088/1361-6382/aabac0;%%

  \bibitem{Wetterich:2003wr} 
  C.~Wetterich,
  ``Gravity from spinors,''
  Phys.\ Rev.\ D {\bf 70}, 105004 (2004)
 % doi:10.1103/PhysRevD.70.105004
  [hep-th/0307145].
  %%CITATION = doi:10.1103/PhysRevD.70.105004;%%

  \bibitem{Ambjorn:2012jv} 
  J.~Ambjorn, A.~Goerlich, J.~Jurkiewicz and R.~Loll,
  ``Nonperturbative Quantum Gravity,''
  Phys.\ Rept.\  {\bf 519}, 127 (2012)
 % doi:10.1016/j.physrep.2012.03.007
  [arXiv:1203.3591 [hep-th]].
  %%CITATION = doi:10.1016/j.physrep.2012.03.007;%%
 
  \bibitem{Markopoulou:2003ps} 
  F.~Markopoulou and L.~Smolin,
  ``Quantum theory from quantum gravity,''
  Phys.\ Rev.\ D {\bf 70}, 124029 (2004)
 % doi:10.1103/PhysRevD.70.124029
  [gr-qc/0311059].
  %%CITATION = doi:10.1103/PhysRevD.70.124029;%%

  \bibitem{Erlich:2018qfc} 
  J.~Erlich,
  ``Stochastic Emergent Quantum Gravity,''
  Class.\ Quant.\ Grav.\  {\bf 35}, no. 24, 245005 (2018)
%  doi:10.1088/1361-6382/aaeb55
  [arXiv:1807.07083 [gr-qc]].
  %%CITATION = doi:10.1088/1361-6382/aaeb55;%%
   
\bibitem{Amati:1981rf} 
  D.~Amati and G.~Veneziano,
  ``Metric From Matter,''
  Phys.\ Lett.\  {\bf 105B}, 358 (1981).
%  doi:10.1016/0370-2693(81)90779-6
  %%CITATION = doi:10.1016/0370-2693(81)90779-6;%%
  
  \bibitem{Amati:1981tu} 
  D.~Amati and G.~Veneziano,
  ``A Unified Gauge and Gravity Theory With Only Matter Fields,''
  Nucl.\ Phys.\ B {\bf 204}, 451 (1982).
%  doi:10.1016/0550-3213(82)90201-2
  %%CITATION = doi:10.1016/0550-3213(82)90201-2;%%

   
 %%%%%%%%%%%%%%%%%%%%%%%%%%%%%%%%%%%%%%%%%%%%%%%%%%%%%%%%%%%%  

\end{thebibliography}
\end{document}